\documentclass[a4paper,12pt,titlepage]{article} 
\usepackage{german, ngerman,makeidx,amssymb,amsmath,amsthm,graphicx}

\begin{document}

\title{{\huge Einsatz von Hilfsmitteln zur Versuchsvorbereitung und Protokollierung im Anf\"angerpraktikum Physik
}\\
\vspace{0.5cm}
{\large Lehrforschungsbericht erstellt im Rahmen einer hochschuldidaktischen Weiterbildung}
\vspace{1cm}
\author{
Dr. Kristin Kliemt
}
\date
{Frankfurt am Main, August 2019\\
kliemt@physik.uni-frankfurt.de}
}
\maketitle

\noindent{\bf Abstract:} The Beginners Lab Course in Physics is scheduled very early in the course of studies not only for students of physics but also for students for which physics is a minor subject. We provide all partipants with several auxilliary materials for the preparation of the experiments and for writing a protocol. In our study, we enlighten the effect of the usage of these materials on the success of passing the lab course.

-----

\noindent{\bf Abstract:} Das Anf\"angerpraktikum Physik ist eine fr\"uh im Studienverlauf stattfindende Veranstaltung sowohl f\"ur Hauptfachstudierende Physik, alsauch f\"ur Studierende mit Nebenfach Physik. Die vorliegende Untersuchung beleuchtet den Effekt der Verwendung von Hilfsmitteln zur Versuchsvorbereitung und Protokollierung auf das erfolgreiche Bestehen des Praktikums.



\section{Forschungsfrage}
Ziel der vorliegenden Studie ist es, die eigene Lehre systematisch nach dem Konzept des Forschenden Lehrens \cite{Spinath} zu untersuchen und die Auswirkungen verschiedener Modifikationen im Lehrbetrieb n\"aher zu beleuchten. Folgende wissenschaftliche Hypothesen \cite{Hussy2010} sollten untersucht werden:
\begin{itemize}
\item 
Die Verwendung der gegebenen Hilfsmittel durch die Mitglieder der Experimentalgruppe f\"uhrt zu einem \glqq besseren Bestehen\grqq\, des Praktikums. 
\glqq Besseres Bestehen\grqq\, bedeutet hierbei, 
dass die Studenten nach dem Praktikum das Gef\"uhl haben, etwas gelernt zu haben 
und ihre F\"ahigkeiten in verschiedenen Bereichen verbessert zu haben.
Wenn die Hilfsmittel 
verwendet werden, wird das Praktikum besser bestanden.
In dieser Untersuchung gibt es mehrere Pr\"adikatoren:
\begin{itemize}
\item eine Video-Aufzeichnung der Einf\"uhrungsveranstaltung
\item eine \glqq Checkliste Versuchsvorbereitung\grqq
\item verschiedene Erkl\"arvideos zu den einzelnen Versuchen (Youtube)
\item und eine \glqq Checkliste Protokoll\grqq.
\end{itemize}
\item 
Durch die gelegentliche Erinnerung der Mitglieder der Experimentalgruppe an die Existenz der Hilfsmittel im Verlauf des Praktikums werden diese h\"aufiger verwendet, was das \glqq Bessere Bestehen\grqq\, unterst\"utzt.
Wenn an die Existenz der Hilfsmittel erinnert wird, 
wird das Praktikum besser bestanden. 
Im Unterschied zur Experimentalgruppe wird in der Kontrollgruppe nicht an die Existenz der Hilfsmittel erinnert.
\end{itemize}



\section{Versuchsplan (Pretest- Posttest Control Group Design)}

Die Untersuchung wurde als Vortest-Nachtest-Design mit Kontrollgruppe ohne Ma\ss nahme durchgef\"uhrt.
Die Teilnehmer der Lehrveranstaltung sind Studierende mit Hauptfach Physik, Studierende mit Nebenfach Physik und Lehramtskandidaten f\"ur Physik an Gymnasien (L3).
Das Praktikum findet in den zwei Gruppen (Experimental- und Kontrollgruppe) an zwei verschiedenen Tagen statt. 
Die Experimentalgruppe besteht aus Studierenden mit Hauptfach Physik und Lehramtskandidaten f\"ur Physik an Gymnasien (L3). Die Kontrollgruppe besteht aus Studierenden mit Hauptfach Physik und aus Studierenden mit Nebenfach Physik.
\newpage
\twocolumn
Der Versuchsplan gestaltet sich wie folgt:
\begin{enumerate}
\item  Vortest (Umfrage) in beiden Gruppen w\"ahrend der Einf\"uhrungsveranstaltung
\item  Ma\ss nahmen: Durchf\"uhrung von mehreren \glqq Erinnerungsumfragen\grqq\, in der Experimentalgruppe w\"ahrend des Semesters, um an die Existenz der Hilfsmittel zu erinnern.
\item Nachtest (Umfrage) in beiden Gruppen am Semesterende. 
\end{enumerate}

\section{Beschreibung der Lehrveranstaltung}
\begin{figure}[htbp]
\includegraphics[width=0.9\columnwidth]{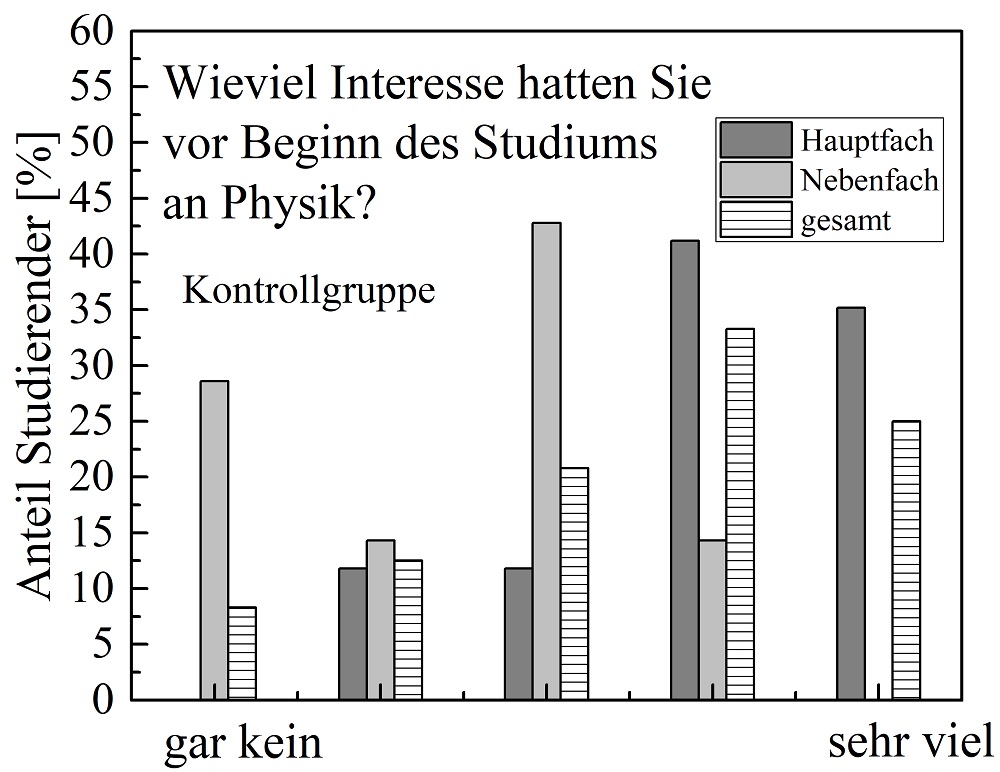}
\caption{Ergebnis der Befragung der Mitglieder der Kontrollgruppe nach ihrem Interesse an Physik vor Beginn des Studiums. Die Kontrollgruppe setzt sich aus Studierenden mit Hauptfach Physik und Studierenden mit Physik als Nebenfach zusammen.}
\label{interesse1}
\end{figure}
\begin{figure}[htbp]
\includegraphics[width=0.9\columnwidth]{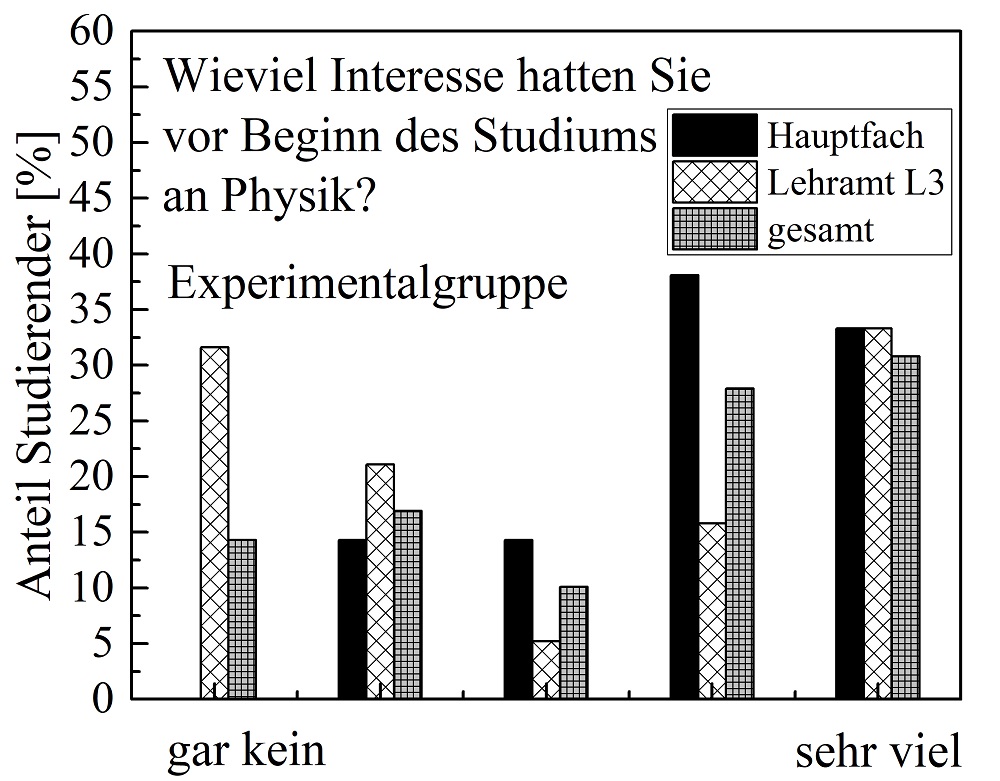}
\caption{Ergebnis der Befragung der Mitglieder der Experimentalgruppe nach ihrem Interesse an Physik vor Beginn des Studiums. Die Experimentalgruppe setzt sich aus Studierenden mit Hauptfach Physik und Lehramtskandidaten Physik (L3) zusammen.}
\label{interesse2}
\end{figure}
Die Untersuchung wird in zwei Studierendengruppen (je ca. 30 Teilnehmer) im \linebreak Anf\"angerpraktikum Physik durchgef\"uhrt. \linebreak Bei den Studierenden handelt es sich um Physikstudenten im 2./3. Semester, Studierende mit Nebenfach Physik (Meteorologie, Mathematik, Informatik) sowie Lehramtskandidaten (L3).
Im Praktikum werden die Versuche in Zweiergruppen durchgef\"uhrt. Jeder Tutor betreut ca. 8 Personen an einem Versuchstag (4 Stunden).
Die Veranstaltung ist derzeit mit 8 Creditpoints dotiert. Neben der Pr\"asenzzeit von 4 h pro Woche ist eine Vor- und Nachbereitungszeit von 12 h pro Woche vorgesehen.

Ein Auszeichnungsmerkmal des Physikers ist die F\"ahigkeit sich selbst\"andig in neue Themen einarbeiten zu k\"onnen sowie bis dato ungel\"oste Problemstellungen selbst\"andig bearbeiten zu k\"onnen. Dementsprechend ist die Ausbildung zu gestalten. Exemplarisch werden diese F\"ahigkeiten im Praktikum mit Hilfestellung eines Tutors ge\"ubt.
Im \linebreak Anf\"anger-Praktikum ist deshalb eine gute Vorbereitung (selbst\"andiges Einarbeiten in ein Thema zu Hause) der Studierenden auf den jeweiligen Versuch unverzichtbar, um w\"ahrend des Praktikums die Aufgaben zum Versuch m\"oglichst selbst\"andig l\"osen zu \linebreak k\"onnen. Im Idealfall w\"urde das bedeuten, dass der Studierende sich in das Thema eingearbeitet hat. W\"ahrend des Eingangs- Kolloquiums werden offene Fragen, die sich \linebreak w\"ahrend der Vorbereitung ergeben haben, diskutiert, sodass der Studierende den Versuch vollkommen selbst\"andig l\"osen kann. In der Praxis wird ein gut vorbereiteter Studierender den Versuch mit wenigen Hilfestellungen des Tutors bearbeiten k\"onnen.
Das kommt dem Idealfall sehr nahe und das Ausbildungsziel kann hier erreicht werden. Bei einem schlecht bis gar nicht vorbereiteten Studierenden ist die selbst\"andige Bearbeitung des Versuches nicht m\"oglich. In diesem Fall macht der Tutor den Versuch und der Student ist nur Zuschauer. Das Ausbildungsziel wird in diesem Fall nicht erreicht. Au\ss erdem werden Tutoren hier unn\"otig gebunden, sodass vorbereitete Studenten insbesondere bei weiterf\"uhrenden Fragen, wie z.B. Einbindung des Versuchs in einen Gesamtzusammenhang, zu kurz kommen. \linebreak Einen wichtigen Teil des Praktikums stellt die Protokollierung der Versuchsergebnisse dar. Die Protokolle zu den Versuchen werden durch die Tutoren korrigiert. Ein Versuch ist erst dann erfolgreich absolviert, wenn das
erstellte Protokoll keine gravierenden \linebreak M\"angel mehr aufweist. Die Anforderungen, die an ein Protokoll gestellt werden, sind in einer \glqq Checkliste zum Schreiben von Protokollen\grqq\, zusammengefasst.

\begin{figure}[htbp]
\includegraphics[width=0.9\columnwidth]{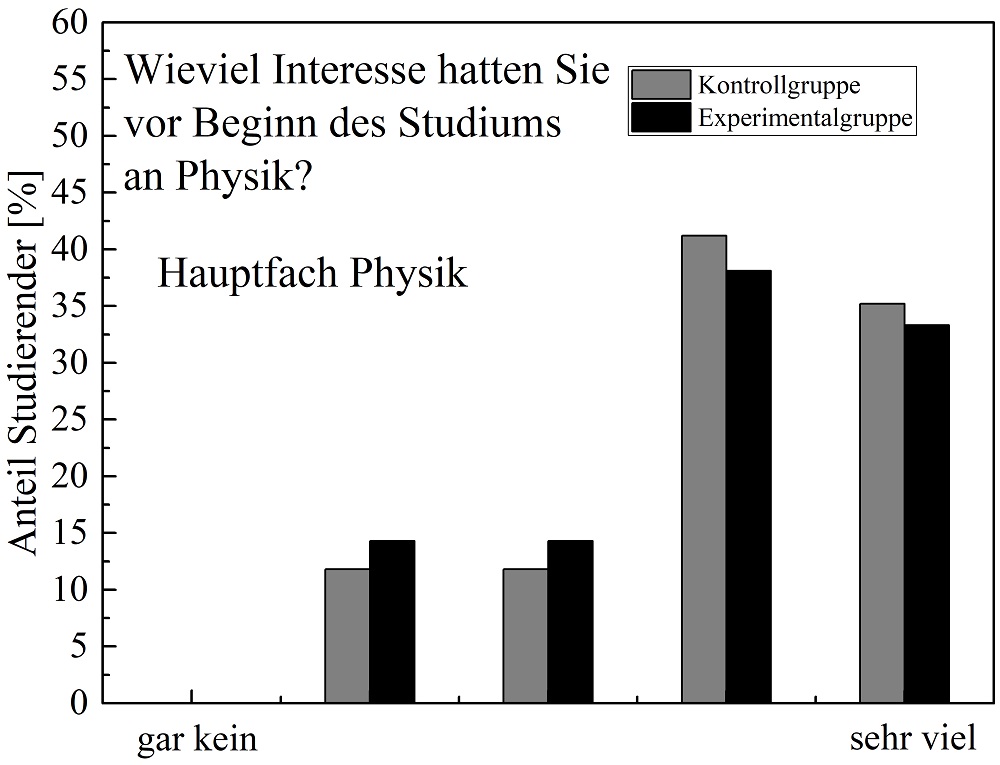}
\caption{Vergleich des Ergebnisses der Befragung der Studierenden mit Hauptfach Physik der Experimental- und der Kontrollgruppe nach ihrem Interesse an Physik vor Beginn des Studiums.}\label{interesse3}
\end{figure}

\begin{figure}[htbp]
\includegraphics[width=0.9\columnwidth]{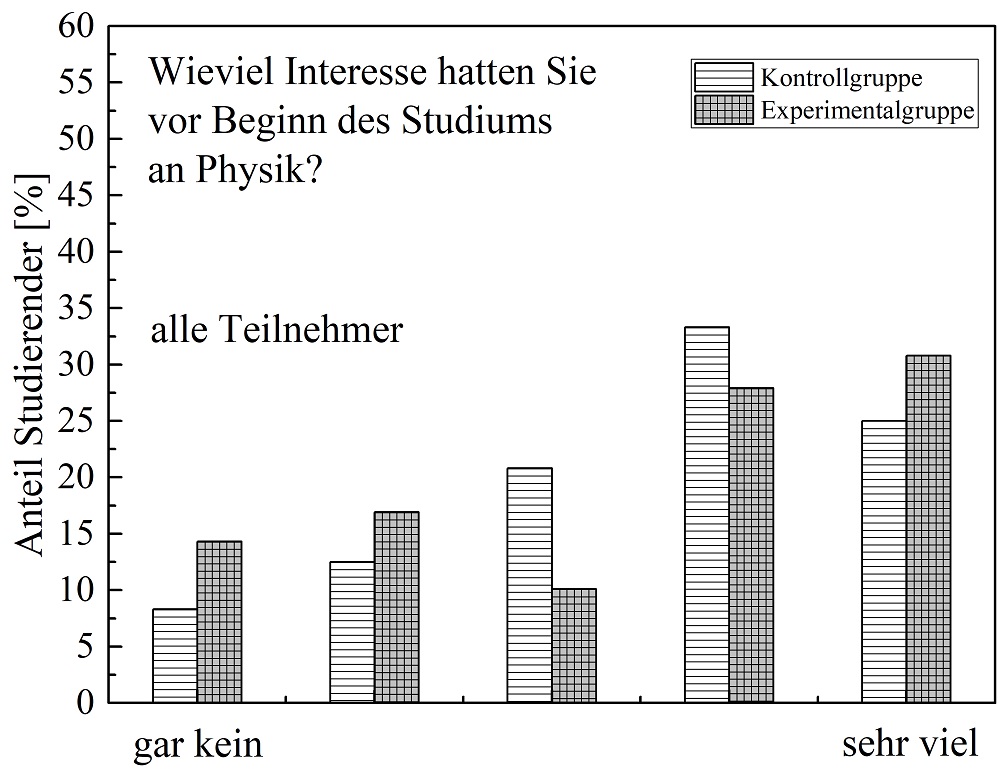}
\caption{Vergleich des Ergebnisses der Befragung aller Studierenden der Experimental- und der Kontrollgruppe nach ihrem Interesse an Physik vor Beginn des Studiums.}\label{interesse4}
\end{figure}

\section{Experimentelle Ergebnisse und Diskussion}

\subsection{Vortest}
Das Physikalische Anf\"angerpraktikum Teil 1 findet f\"ur Studierende mit Hauptfach Physik turnusgem\"a\ss\, im zweiten und dritten Semester statt. Die Erfahrungen der Teilnehmer bez\"uglich des Faches Physik sind \"ublicherweise w\"ahrend dieser Zeit noch stark von den Erlebnissen und der erlangten Motivation w\"ahrend der Schulzeit gepr\"agt. Der individuelle Kenntnisstand und die jeweiligen fachlichen F\"ahigkeiten h\"angen stark von der schulischen Vorbildung bzw. wahrgenommen mathematisch/physikalischen Zusatzangeboten ab. 
Zu Beginn der Veranstaltung wurde mittels einer Umfrage, die sowohl in der Kontroll- als auch in der Experimentalgrupppe durchgef\"uhrt wurde, das Interesse der Teilnehmer an Physik sowie die besuchten Leistungskurse in der Schule ermittelt. Ausserdem wurden die Teilnehmer gebeten, Ihre physikalische Vorbildung als Vorbereitung auf das Studium, ihre Erfahrung im Durchf\"uhren von Experimenten sowie im \linebreak Schreiben von Protokollen selbst einzusch\"at- zen. Die Angaben f\"ur die Kontrollgruppe werden immer in Klammern angegeben.

\begin{figure}[htbp]
\includegraphics[width=0.9\columnwidth]{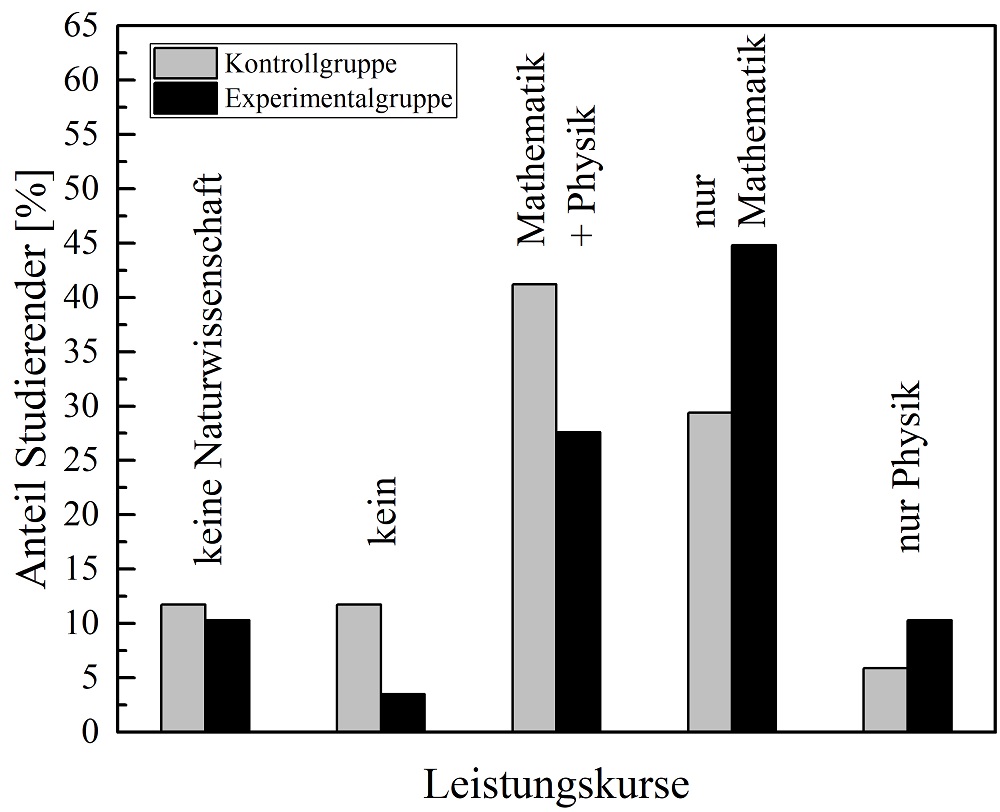}
\caption{Angaben der Teilnehmer aus Experimental- und Kontrollgruppe zu den jeweils besuchten Leistungskursen.}\label{LK}
\end{figure}

\begin{figure}[htbp]
\includegraphics[width=0.9\columnwidth]{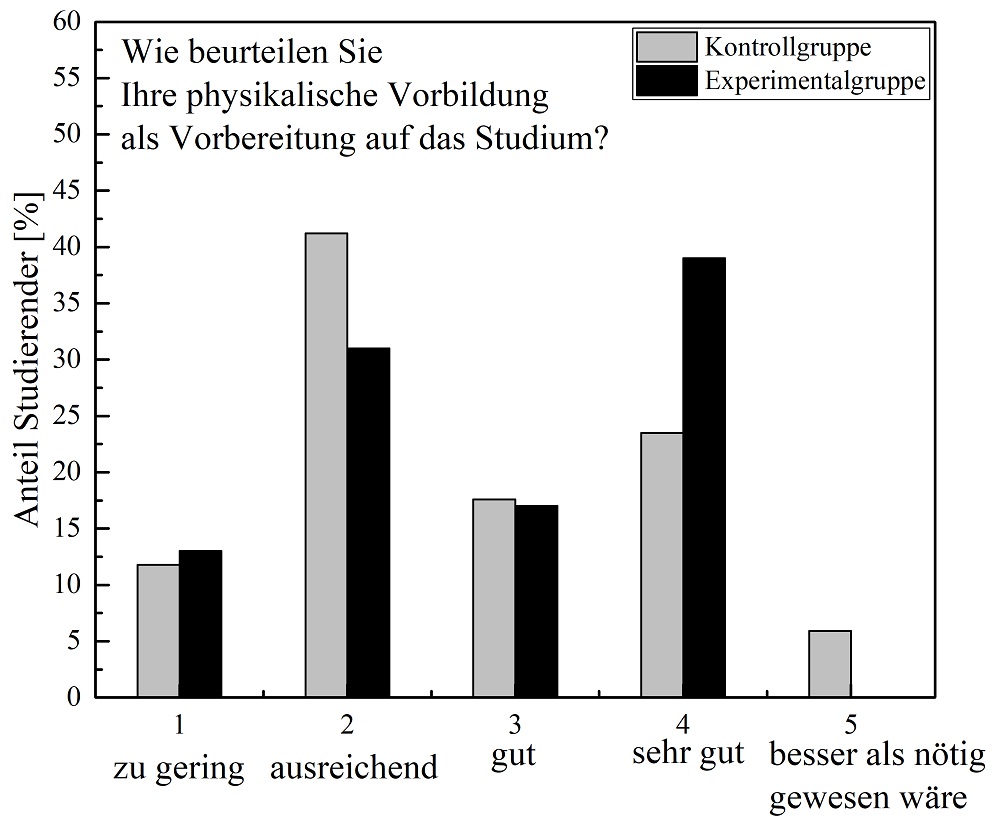}
\caption{Einsch\"atzung der Teilnehmer aus Experimental- und Kontrollgruppe zu ihrer eignen physikalischen Vorbildung in Vorbereitung auf das Studium.}\label{Vorbildung}
\end{figure}

\subsection*{Interesse an Physik}

Studierende der Kontrollgruppe, bestehend aus Studierenden mit Hauptfach Physik und Nebenfachstudierenden, antworteten auf die Frage \glqq Wieviel Interesse hatten Sie vor Beginn des Studiums an Physik?\grqq\, (Abb.~\ref{interesse1}). W\"ahrend Physikstudierende im Hauptfach erwartungsgem\"a\ss\, ein hohes fachliches Interesse angaben, antworteten fast 30\% der Nebenfachstudierenden,\linebreak dass sie vor Beginn des Studiums gar kein Interesse an Physik hatten.
In der Experimentalgruppe, bestehend aus Studierenden mit Hauptfach Physik und Kandidaten f\"ur Lehramt Physik an Gymnasien (L3), ergab sich ein \"ahnliches Bild (Abb.~\ref{interesse2}). W\"ahrend das Interesse am Fach bei den Hauptfachstudierenden mehrheitlich als gro\ss\, einge-\linebreak sch\"atzt wurde, zeigte sich im Gegensatz dazu bei den Lehramtskandidaten gro\ss e Heterogenit\"at. Ein Drittel dieser Teilnehmer gab an, das Studium mit sehr viel Interesse an Physik begonnen zu haben, w\"ahrend ein weiteres Drittel dies ohne jegliches Interesse am Fach tat. Als 
Ursache f\"ur die Abweichung von der Normalverteilung kann hier davon ausgegangen werden, dass hier ein nicht normalverteiltes Merkmal \linebreak abgefragt wurde (\cite{Moosbrugger2007}, S.93).

\begin{figure}[htbp]
\centering
\includegraphics[width=0.9\columnwidth]{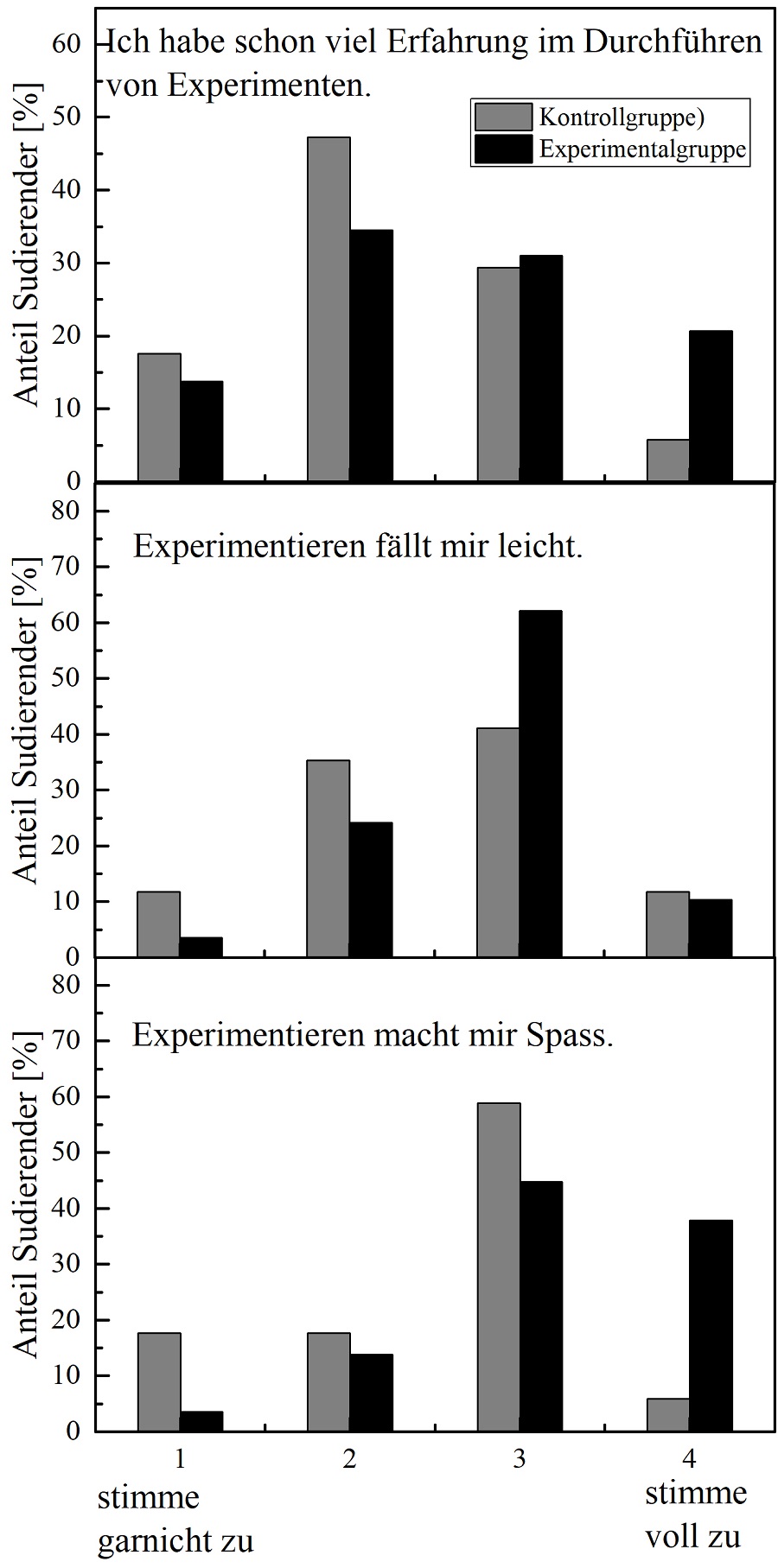}
\caption{Vergleich der Teilnehmer der Experimental- und der Kontrollgruppe. {\it Oben:} Selbsteinsch\"atzung der Erfahrung im Durchf\"uhren von Experimenten; {\it Mitte:} Selbsteinsch\"atzung zur Aussage: \glqq Experimentieren f\"allt mir leicht\grqq\, ; {\it Unten:} Selbsteinsch\"atzung zur Aussage: \glqq Experimentieren macht mir Spa\ss \grqq\, .}
\label{ExpAnfang}
\end{figure}

Studierende im Hauptfach Physik antworteten auf die Frage \glqq Wieviel Interesse hatten Sie vor Beginn des Studiums an Physik?\grqq. Auf einer 5-stufigen Skala (1 = gar kein Interesse, 5 = sehr viel Interesse) wurde das Interesse von 71,4 \% der Teilnehmer der Experimentalgruppe (76,4 \% Kontrollgruppe) mit 4 oder 5 bewertet (Abb.~\ref{interesse3}). Damit gaben sie im Mittel die Bewertung $3,9$ ($4,0$) ab. Betrachtet man die Gesamtheit der Teilnehmer incl. Studierende mit Nebenfach Physik und Lehramtskandidaten L3 sinkt der Anteil der Studierenden mit gro\ss em Interesse (Stufe 4 oder 5) auf $58,7 \%$ der Teilnehmer (58,3 \%), siehe Abb.~\ref{interesse4}. 
Insgesamt gesehen kann man von einer \"ahnlichen Interessenlage in Experimental- und Kontrollgruppe in Bezug auf fachliche Inhalte ausgehen.

\subsection*{Leistungskurse und Vorbildung}
Alle Teilnehmer wurden nach den besuchten Leistungskursen in der Schule befragt. 82,7\% (76,5\%)
der Teilnehmer gaben an, \linebreak einen Mathematik- oder Physikleistungskurs besucht zu haben, Abb.~\ref{LK}. Damit haben die Teilnehmer aus Experimental- und Kontrollgruppe \"ahnliche Voraussetzungen in Bezug auf fachliche Kenntnisse und F\"ahigkeiten.
Die Studierenden antworteten auf die Frage \glqq Wie beurteilen Sie Ihre physikalische Vorbildung als Vorbereitung auf das Studium?\grqq , (Abb.~\ref{Vorbildung}).  Auf einer 5-stufigen Skala (1 = zu gering, 2 = ausreichend, 3 = gut, 4 = sehr gut, 5 = besser als n\"otig gewesen w\"are) gaben 44\% (53\%) der Studierenden eine Bewertung der Stufe 1 oder 2 ab und sch\"atzten damit ihre Vorbildung als \glqq zu gering\grqq\, oder grade als \glqq ausreichend\grqq\, ein. Dem subjektiven Empfinden nach f\"uhlen sich damit die Teilnehmer der Kontrollgruppe durch die \linebreak schulische Ausbildung sogar noch etwas \linebreak schlechter auf das Studium vorbereitet als die Teilnehmer der Experimentalgruppe.

\begin{figure}
\centering
\includegraphics[width=0.9\columnwidth]{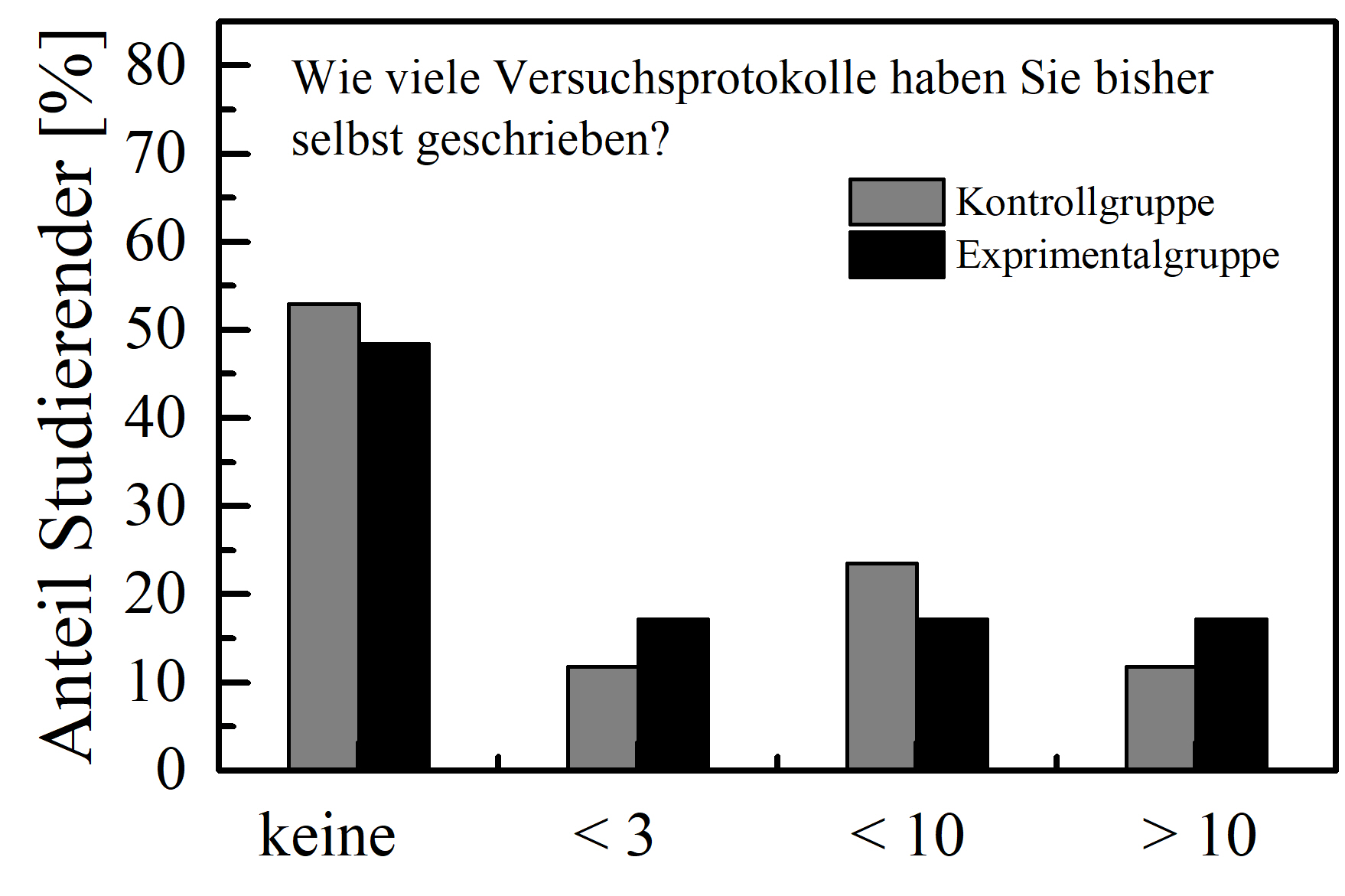}
\caption{Vergleich der Teilnehmer der Experimental- und der Kontrollgruppe bzgl. der Anzahl bisher selbst geschriebener Protokolle.}
\label{Anzahl}
\end{figure}

\subsection*{Durchf\"uhren von Experimenten}
Die Teilnehmer in der Experimentalgruppe sch\"atzten auf einer 4-stufigen Skala (1 = stimme gar nicht zu, 4 = stimme voll zu) ihre Erfahrung im Durchf\"uhren von Experimenten (Abb.~\ref{ExpAnfang}, oben) mit einem Mittelwert der Stufe $2,6$ \"ahnlich ein wie die Teilnehmer der Kontrollgruppe (Stufe $2,2$). Sie stimmten auch der Aussage \glqq Experimentieren f\"allt mir leicht\grqq\, (Abb.~\ref{ExpAnfang}, mitte) mit Stufe $2,8$ als Mittelwert etwas h\"aufiger zu als die Teilnehmer der Kontrollgruppe (Stufe $2,5$). Der Aussage \glqq Experimentieren macht mir Spass\grqq\, (Abb.~\ref{ExpAnfang}, unten) wurde in der Experimentalgruppe, Mittelwert Stufe $3,2$, deutlich h\"aufiger zugestimmt als in der Kontrollgruppe (Stufe $2,5$).

\begin{figure}
\centering
\includegraphics[width=0.9\columnwidth]{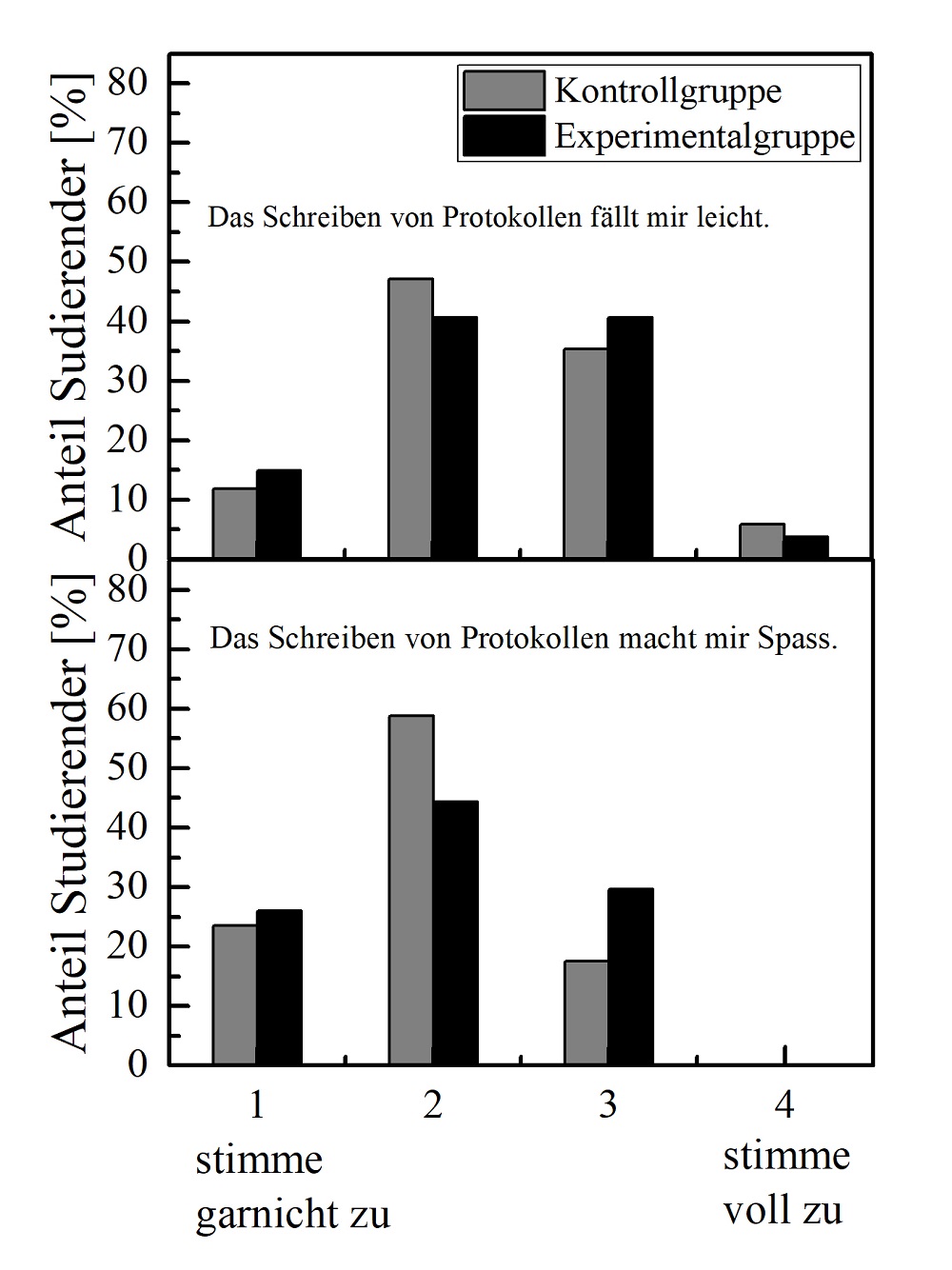}
\caption{Vergleich der Teilnehmer der Experimental- und der Kontrollgruppe. {\it Oben:} Selbsteinsch\"atzung zur Aussage: \glqq Das Schreiben von Protokollen f\"allt mir leicht\grqq ; {\it Unten:} Selbsteinsch\"atzung zur Aussage: \glqq Das Schreiben von Protokollen macht mir Spa\ss \grqq .}
\label{ProtokollLeichtAnfang}
\end{figure}
\subsection*{Schreiben von Protokollen}

48,4 \% (52,9 \%) der Teilnehmer gaben an, bisher keine Erfahrung im Schreiben von Protokollen zu haben (Abb.~\ref{Anzahl}). Auf einer 4-stufigen Skala (1 = stimme gar nicht zu, 4 = stimme voll zu) wurde der Aussage \glqq Das Schreiben von Protokollen f\"allt mir leicht.\grqq\,  nur von 3,7 \% (5,8 \%) der Teilnehmer voll zugestimmt (Abb.~\ref{ProtokollLeichtAnfang}, oben). Der Mittelwert lag hier bei Stufe $2,33$ ($2.35$). Der Aussage \glqq Das Schreiben von Protokollen macht mir Spass.\grqq\, wurde von keinem Teilnehmer voll zugestimmt. Der Mittelwert lag hier bei Stufe $2,04$ ($1,94$) (Abb.~\ref{ProtokollLeichtAnfang}, unten).

\subsection*{Kenntnis der Hilfsmittel}
\begin{figure}[htbp]
\centering
\includegraphics[width=0.9\columnwidth]{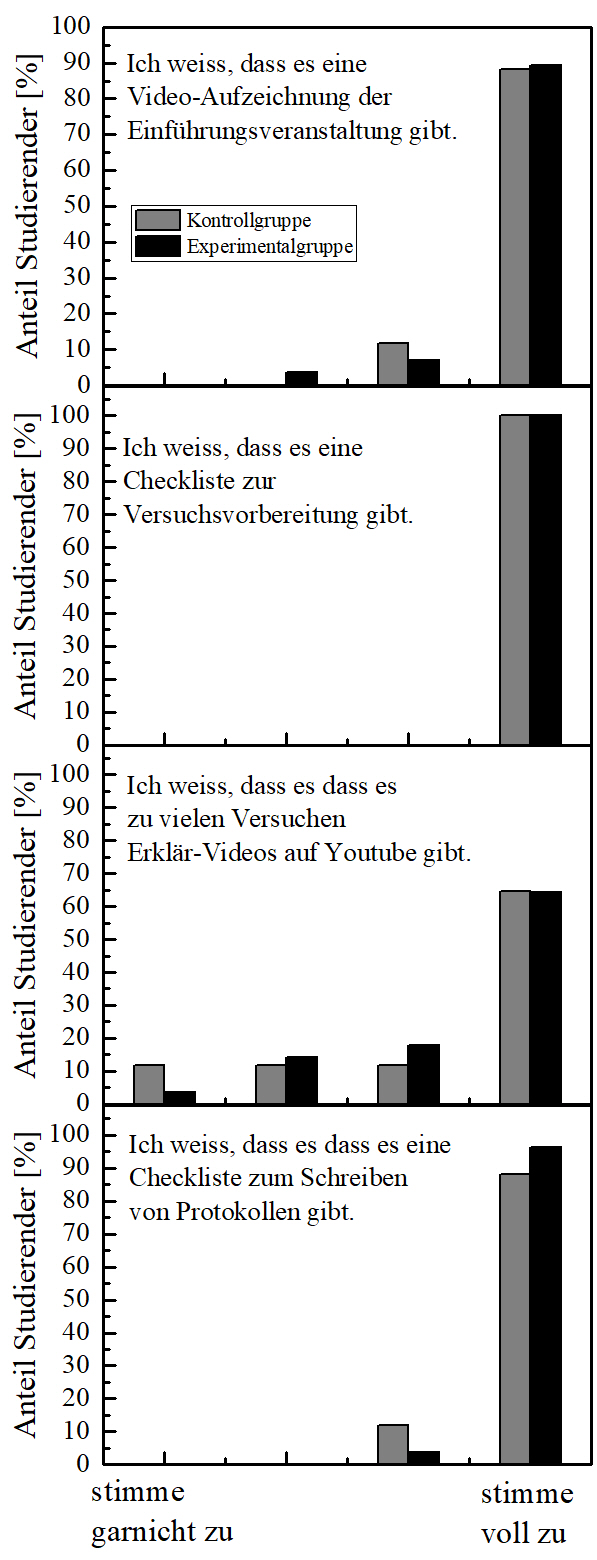}
\caption{Vergleich der Teilnehmer der Experimental- und der Kontrollgruppe: Selbstauskunft zur Kenntnis der Hilfsmittel.}
\label{Checklisten}
\end{figure}

Die verschiedenen, neben den Versuchsanleitungen existierenden Hilfsmittel, wurden den Studierenden in der Einf\"uhrungsveranstaltung vorgestellt. Am Ende derselben Veranstaltung wurden die Studierenden im Vortest auch zur Kenntnis der einzelnen Hilfsmittel befragt (Abb.~\ref{Checklisten}).
Die Videoaufzeichnung der Einf\"uhrungsveranstaltung kennen zu diesem Zeitpunkt $89,3\%$ \linebreak ($88,2\%$) der Teilnehmer, die 
\glqq Checkliste Vorbereitung\grqq\, sogar 100\% (100\%) und auch die zur Verf\"ugung stehende \glqq Checkliste Protokoll\grqq\,  wird von einem hohen Anteil der Studierenden gekannt 96,4\% (88,2\%). 
Bei den Videos zu den einzelnen Versuchen gaben lediglich 64,3\% (64,6\%) der Studierenden an, diese zu kennen.

\subsection*{Vergleich von Experimental- und Kontrollgruppe nach dem Vortest}
Der Vergleich beider Gruppen zeigte, dass man insgesamt gesehen von einer \"ahnlichen Interessenlage in Bezug auf fachliche Inhalte zu Beginn des Kurses ausgehen kann. Ebenso haben die Teilnehmer beider Gruppen \"ahnliche Voraussetzungen in Bezug auf fachliche Kenntnisse und F\"ahigkeiten und die Vorbildung, was die Art und Anzahl der besuchten Leistungskurse in der Schule betrifft. Allerdings f\"uhlt sich ein grosser Anteil der Teilnehmer beider Gruppen dem subjektiven Empfinden nach durch die schulische Ausbildung unzureichend auf das Studium vorbereitet. Hier war der Anteil in der Kontrollgruppe sogar noch etwas h\"oher als in der Experimentalgruppe.
Die Erfahrung im Experimentieren der Teilnehmer beider \linebreak Gruppen wurde \"ahnlich eingesch\"atzt. Ebenso wurde der Aussage \glqq Experimentieren f\"allt mir leicht.\grqq\, \"ahnlich h\"aufig zugestimmt.
 Der Aussage \glqq Experimentieren macht mir Spa\ss .\grqq\, wurde in der Experimentalgruppe h\"aufiger zugestimmt als in der Kontrollgruppe. Die Erfahrung im Schreiben von Protokollen der Teilnehmer beider Gruppen wurde \"ahnlich eingesch\"atzt. Ebenso wurde den Aussagen \glqq Das Schreiben von Protokollen f\"allt mir \linebreak leicht.\grqq\, und \glqq Das Schreiben von Protokollen macht mir Spa\ss .\grqq\, \"ahnlich h\"aufig zugestimmt. Der Kenntnisstand bzgl. der zur \linebreak Verf\"ugung stehenden Hilfsmittel war in beiden Gruppen nach der Einf\"uhrungsveranstaltung \"ahnlich.

\subsection{Ma\ss nahme}
Den Teilnehmern beider Gruppen standen w\"ahrend des Praktikums alle Hilfsmittel zur Verf\"ugung.
Die Hilfsmittel wurden in einer Einf\"uhrungsverstaltung vor Beginn des Praktikums allen Teilnehmern vorgestellt.
In der Experimentalgruppe wurde w\"ahrend des Semesters mehrfach nach der Benutzung der Hilfsmittel gefragt und somit an deren Existenz erinnert. Diese Ma\ss nahme wurde in der Kontrollgruppe nicht durchgef\"uhrt.

\subsection{Nachtest}
Im Nachtest wurde in der Umfrage 
keine neutrale Mittelkategorie angeboten um der \glqq Tendenz zur Mitte\grqq\, (\cite{Moosbrugger2007}, S.60) entgegenzuwirken.
\subsection*{Experimentieren und Protokollieren}

\begin{figure}[htbp]
\includegraphics[width=0.9\columnwidth]{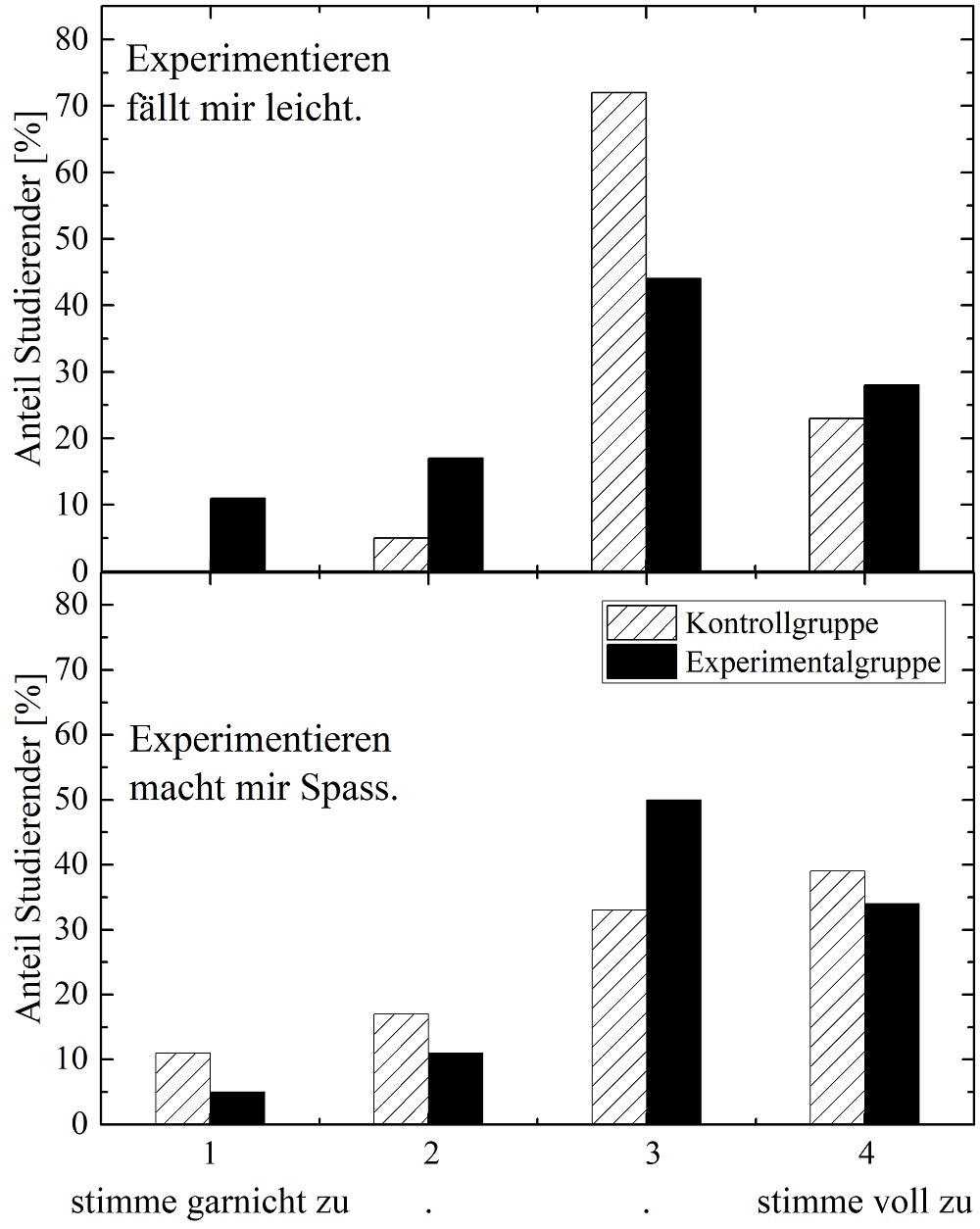}
\caption{Vergleich der Aussagen der Teilnehmer der Experimental- und der Kontrollgruppe im Nachtest. {\it Oben:} Selbsteinsch\"atzung zur Aussage: \glqq Experimentieren f\"allt mir leicht\grqq ; {\it Unten:} Selbsteinsch\"atzung zur Aussage: \glqq Experimentieren macht mir Spa\ss \grqq .}
\label{expEnde}
\end{figure}
Die Teilnehmer in der Experimentalgruppe stimmten auf einer 4-stufigen Skala (1 = stimme gar nicht zu, 4 = stimme voll zu) der Aussage \glqq Experimentieren f\"allt mir leicht\grqq\, (Abb.~\ref{expEnde}, oben) mit Stufe 2,9 (Vortest $2,8$) als Mittelwert etwas weniger h\"aufig zu wie die Teilnehmer der Kontrollgruppe 3,2 (Vortest Stufe $2,5$). Im Vergleich zum Vortest stimmen die Teilnehmer beider Gruppen der Aussage h\"aufiger zu.
Der Aussage \glqq Experimentieren macht mir Spa\ss \grqq\, (Abb.~\ref{expEnde}, unten) wurde in der Experimentalgruppe, Mittelwert Stufe 3,0 (Vortest $3,2$), \"ahnlich\linebreak h\"aufig zugestimmt wie in der Kontrollgruppe 3,1 (Vortest Stufe $2,5$). 
Im Vergleich zum Vortest konnte in der Kontrollgruppe der Spa\ss\, am Experimentieren durch die im Praktikum gesammelten Erfahrungen deutlich \linebreak erh\"oht werden.


\begin{figure}[htbp]
\includegraphics[width=0.9\columnwidth]{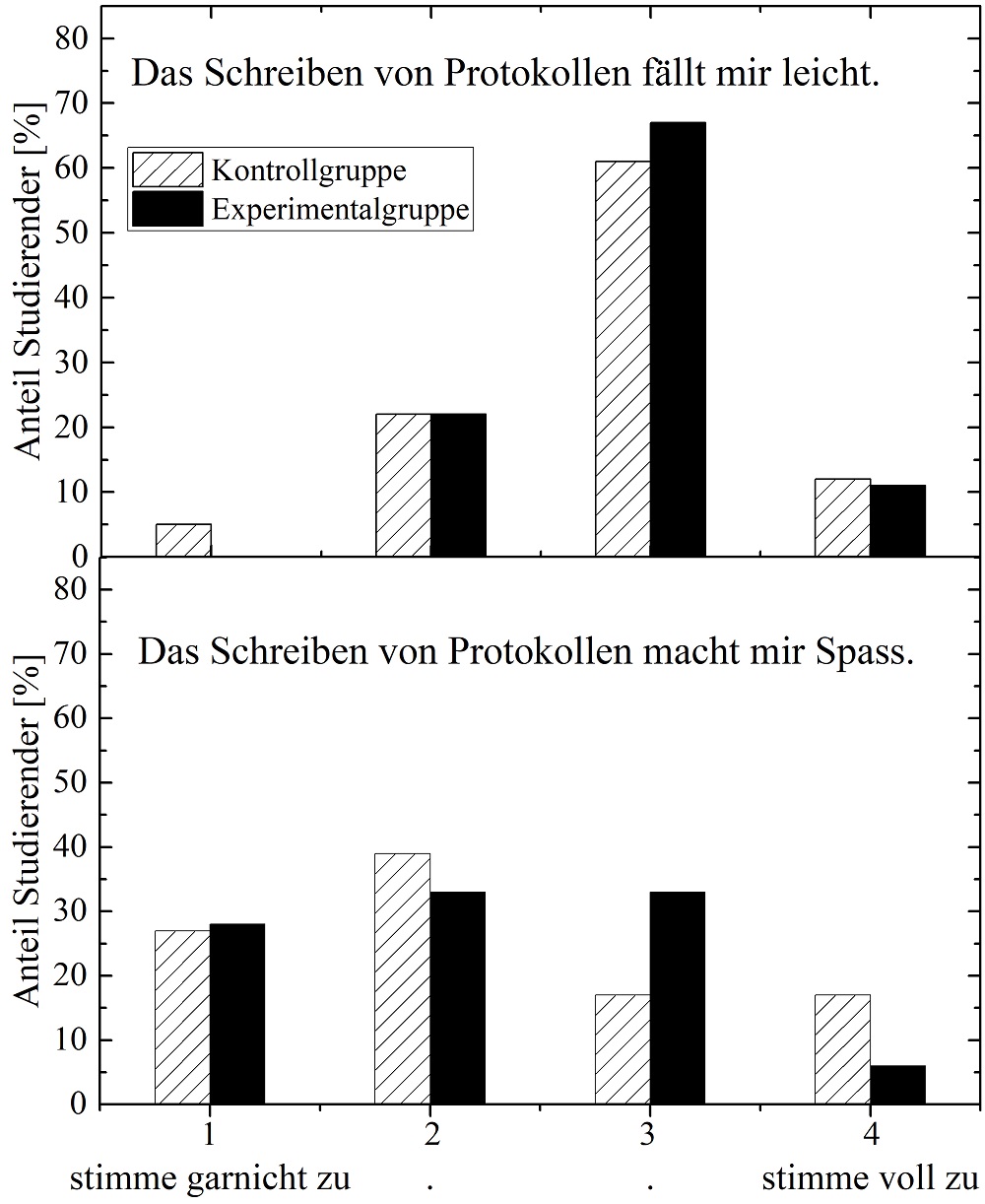}
\caption{Vergleich der Aussagen der Teilnehmer der Experimental- und der Kontrollgruppe im Nachtest. {\it Oben:} Selbsteinsch\"atzung zur Aussage: \glqq Das Schreiben von Protokollen f\"allt mir leicht\grqq\, ; {\it Unten:} Selbsteinsch\"atzung zur Aussage: \glqq Das Schreiben von Protokollen macht mir Spa\ss \grqq\, .}
\label{ProtEnde}
\end{figure}

Auf einer 4-stufigen Skala (1 = stimme gar nicht zu, 4 = stimme voll zu) wurde der Aussage \glqq Das Schreiben von Protokollen f\"allt mir leicht.\grqq\, auch im Nachtest  nur von  11\% (12\%) der Teilnehmer voll zugestimmt (Abb.~\ref{ProtEnde}, oben) (Vortest 3,7 \% (5,8 \%)). Der Mittelwert lag hier bei 2,9 (2,8)\linebreak (Vortest Stufe $2,3$ ($2,4$)). 
Hier ist ein deutlicher Trend zu beobachten: Durch die im Praktikum erworbene Routine, wird diese Aufgabe in beiden Gruppen nun als leichter eingesch\"atzt.
Der Aussage \glqq Das Schreiben von Protokollen macht mir Spa\ss .\grqq\, wurde im Vortest von keinem Teilnehmer voll zugestimmt. 
Im Nachtest stimmten immerhin 6\% (17\%) dieser Aussage voll zu.
Auch bei dieser Aussage f\"uhren die im Praktikum gemachten Erfahrungen zu einer deutlichen Verschiebung des Befragungsergebnisses zu h\"oheren Werten. Der Mittelwert lag hier bei 2,2 (2,2) (Vortest Stufe $2,0$ ($1,9$)) (Abb.~\ref{ProtEnde}, unten).


\subsection*{Benutzung der Hilfsmittel}
\begin{figure}[htbp]
\includegraphics[width=0.9\columnwidth]{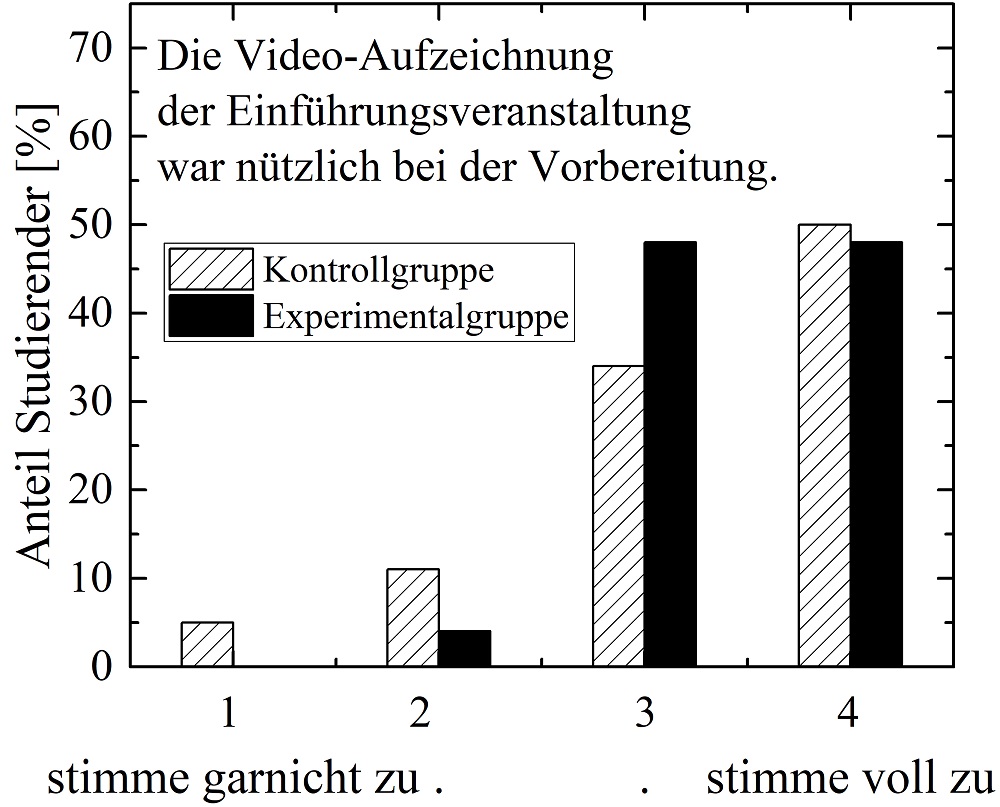}
\caption{Vergleich der Aussagen der Teilnehmer der Experimental- und der Kontrollgruppe im Nachtest bzgl. der Antwort auf die Frage \glqq Die Video-Aufzeichnung der Einf\"uhrungsveranstaltung war n\"utzlich bei der Vorbereitung\grqq .}
\label{EinfEnde}
\end{figure}
Auf einer 4-stufigen Skala (1 = stimme gar nicht zu, 4 = stimme voll zu) wurde der Aussage \glqq Die Video-Aufzeichnung der Einf\"uhrungsveranstaltung war n\"utzlich bei der Vorbereitung.\grqq\, von den Mitgliedern der Experimentalgruppe mit einem Mittelwert von 3,4 etwas h\"aufiger zugestimmt als von denen der Kontrollgruppe (Mittelwert 3,3).

\begin{figure}[htbp]
\includegraphics[width=0.9\columnwidth]{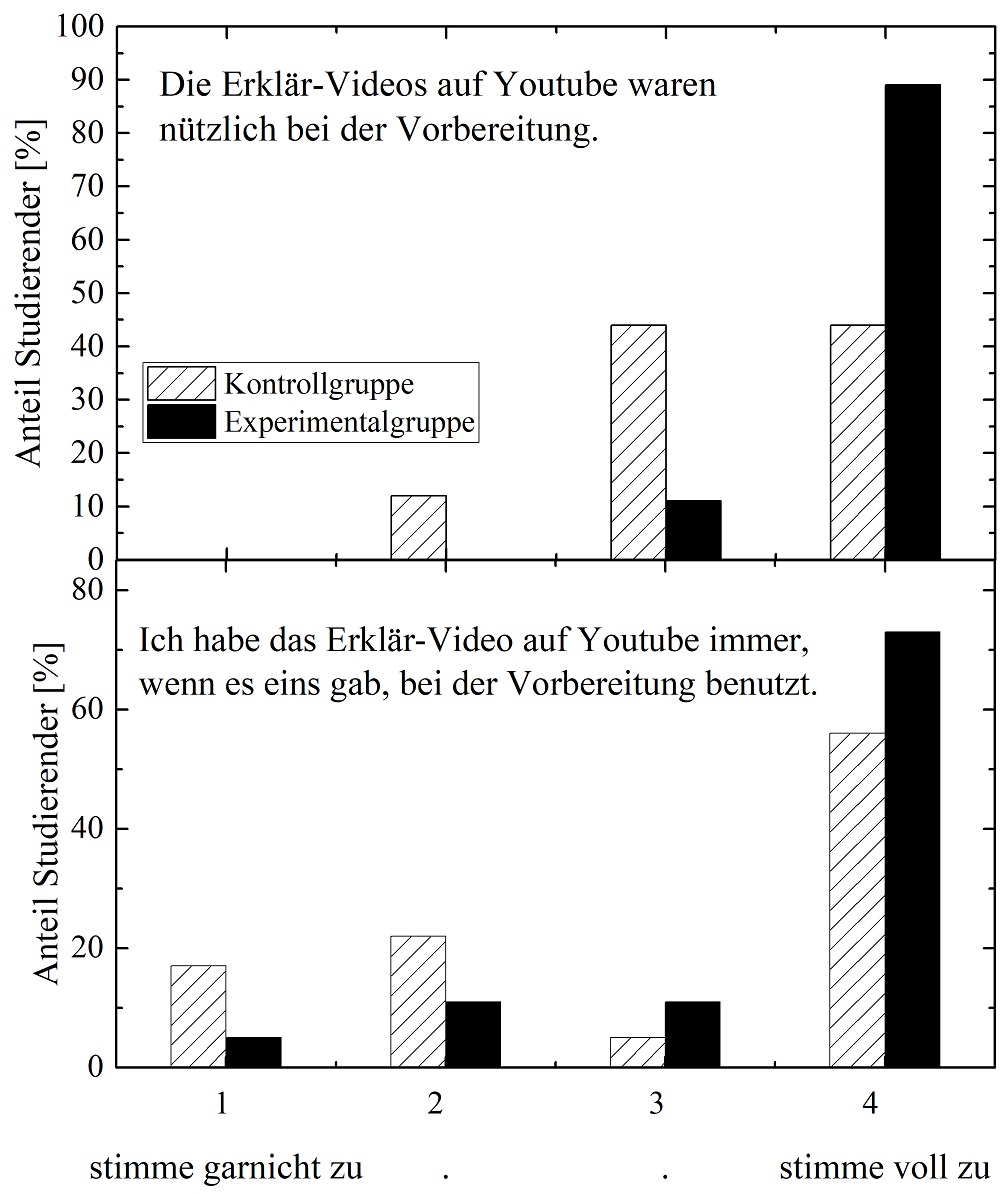}
\caption{Vergleich der Aussagen der Teilnehmer der Experimental- und der Kontrollgruppe im Nachtest bzgl. der Antwort auf die Fragen \glqq Die Erkl\"ar-Videos auf Youtube waren n\"utzlich bei der Vorbereitung.\grqq\, und \glqq Ich habe das Erkl\"ar-Video auf Youtube immer, wenn es eins gab, bei der Vorbereitung benutzt.\grqq }
\label{VideoEnde}
\end{figure}
Auch den Aussagen \glqq Die Erkl\"ar-Videos auf Youtube waren n\"utzlich bei der Vorbereitung.\grqq\, und \glqq Ich habe das Erkl\"ar-Video auf Youtube immer, wenn es eins gab, bei der Vorbereitung benutzt.\grqq\, wurden von den Teilnehmern der Experimentalgruppe mit \linebreak Mittelwert 3,9 
bzw. Mittelwert 3,5 
h\"aufiger zugestimmt als von den Mitgliedern der Kontrollgruppe 3,3 bzw. 3,0.\\
\begin{figure}
\includegraphics[width=0.9\columnwidth]{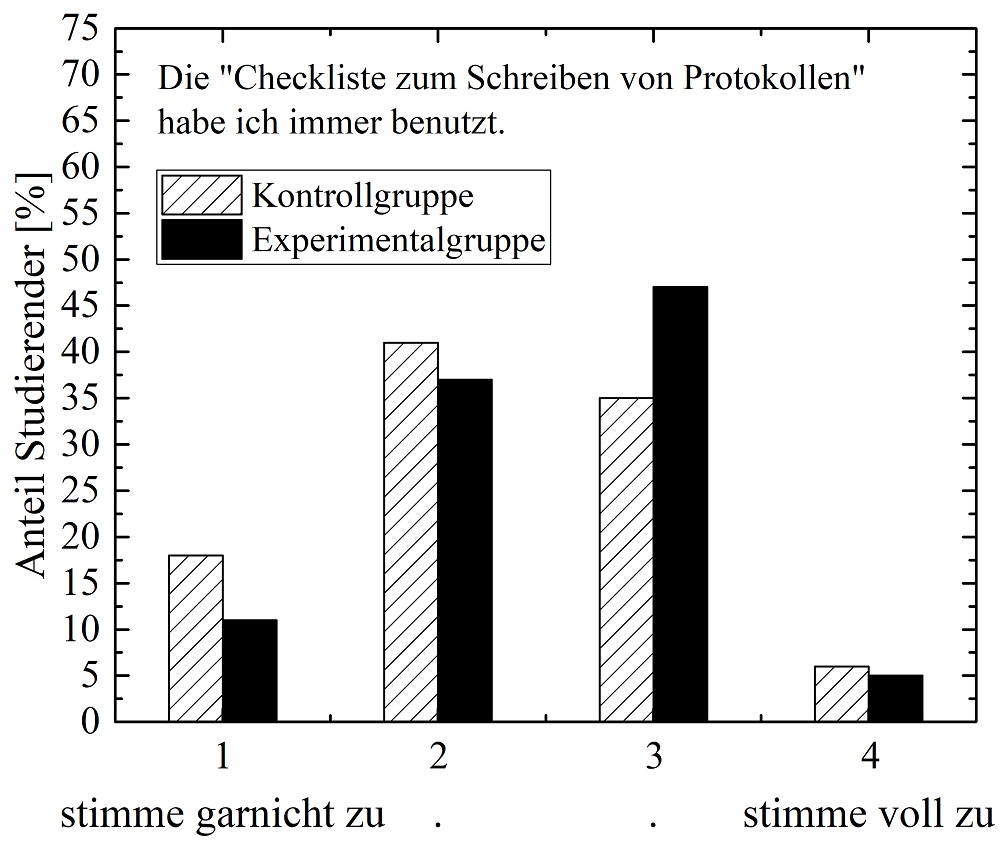}
\caption{Vergleich der Aussagen der Teilnehmer der Experimental- und der Kontrollgruppe im Nachtest bzgl. der Antwort auf die Frage \glqq Die \glqq Checkliste zum Schreiben von Protokollen\grqq\, habe ich immer benutzt.\grqq }
\label{VideoEnde}
\end{figure}
Gleiches gilt f\"ur die Aussage \glqq Die \glqq Checkliste zum Schreiben von Protokollen\grqq\, habe ich immer benutzt.\grqq\, Auch hier stimmten die Mitglieder der Experimentalgruppe mit Mittelwert 2,5 h\"aufger zu, als die Mitglieder der Kontrollgruppe mit Mittelwert 2,2.
\begin{figure}
\includegraphics[width=0.9\columnwidth]{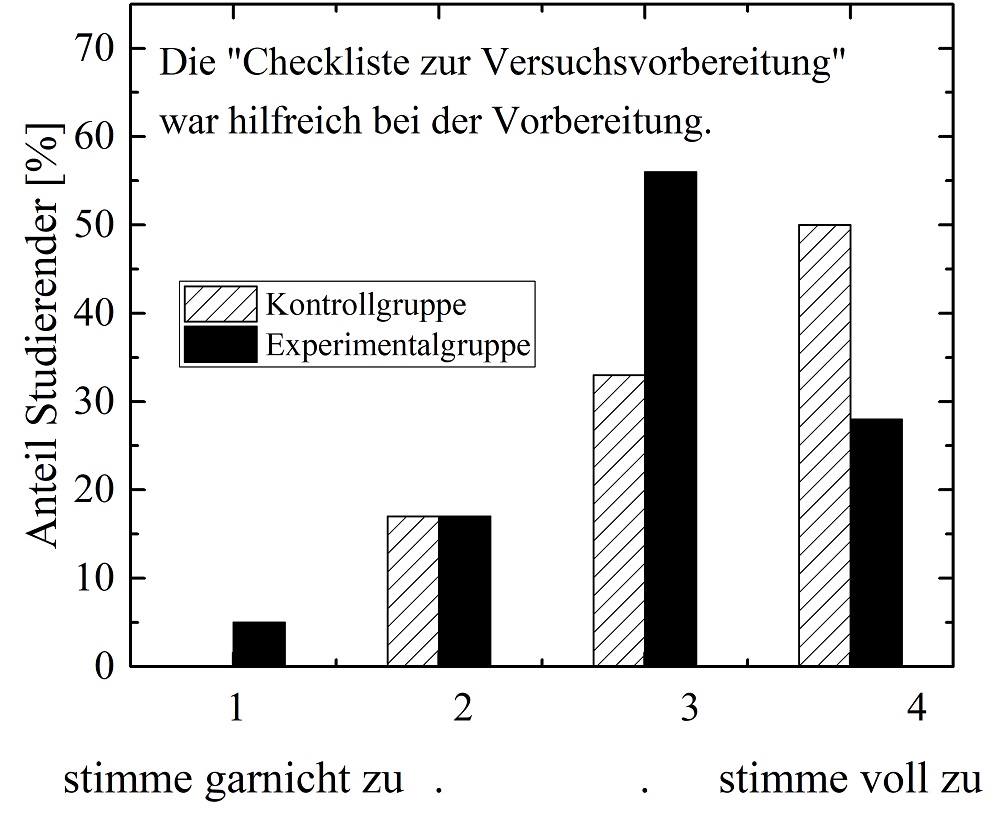}
\caption{Vergleich der Aussagen der Teilnehmer der Experimental- und der Kontrollgruppe im Nachtest bzgl. der Antwort auf die Frage \glqq Die \glqq Checkliste zur Versuchsvorbereitung\grqq\, war hilfreich bei der Vorbereitung.\grqq }
\label{VorbEnde}
\end{figure}
Eine in diesem Semester erstmalig eingesetzte \glqq Checkliste zur Versuchsvorbereitung\grqq\, wurde von den meisten Teilnehmern beider Gruppen
mit Mittelwert 3 (3,3) als hilfreich bei der Vorbereitung angesehen.

\subsection{Zusammenfassung}
In der vorliegenden Untersuchung wurde der Frage nachgegangen, ob die Verwendung der gegebenen Hilfsmittel zu einem \glqq besseren Bestehen\grqq\, des Anf\"anger-Praktikums Physik \linebreak f\"uhrt. Unter \glqq Besserem Bestehen\grqq\, wird hierbei verstanden, dass die Studenten nach dem Praktikum das Gef\"uhl haben, etwas gelernt zu haben und ihre F\"ahigkeiten in verschiedenen Bereichen verbessert zu haben. Das Resultat wurde mit Hilfe einer Selbstein-\linebreak sch\"atzung ermittelt. Au\ss erdem wurde durch Vergleich mit einer Kontrollgruppe untersucht, ob die Erinnerung an die Existenz der Hilfsmittel das Resultat positiv beeinflusst. Die Untersuchung ergab, dass
von den Mitgliedern der Experimentalgruppe, die an die Existenz der Hilfsmittel immer wieder erinnert wurden, die Video-Aufzeichnung der Einf\"uhrungsveranstaltung sowie die Erkl\"ar- \linebreak videos auf Youtube als n\"utzlicher bei der Vorbereitung angesehen wurden als von den Mitgliedern der Kontrollgruppe. Die Videos und die \glqq Checkliste zum Schreiben von Protokollen\grqq\, wurden in der Experimentalgruppe auch h\"aufiger verwendet. 
Eine deutliche Erh\"ohung der Zustimmung ist auch auf die Frage, ob das Schreiben von Protokollen \linebreak leicht f\"allt in beiden Gruppen zu beobachten, wobei auch hier die Zunahme der Zustimmung in der Experimentalgruppe gr\"o\ss er ist. Im Nachtest wurde auch der Frage, ob das Schreiben von Protokollen Spa\ss\, macht in beiden Gruppen deutlich h\"aufiger zugestimmt als im Vortest. 
Die Untersuchung liefert damit deutliche Hinweise darauf, dass das gelegentliche Erinnern an die Existenz der vorhandenen Hilfsmittel zu deren h\"aufigeren Verwendung f\"uhrt und einen positiven Einfluss auf das \glqq Bessere Bestehen\grqq\, des Praktikums hat.

\subsection*{Danksagung}
Ich danke Prof. Dr. C.~Krellner, dem Leiter des Physikalischen Anf\"angerpraktikums Teil 1,  f\"ur seine Unterst\"utzung bei der \linebreak Durchf\"uhrung meiner Untersuchung. Ebenso danke ich den Tutoren D.-M.~Tran, \linebreak S.~Rongstock und E.~Lorenz f\"ur ihre Mithilfe.  S.~Bergmann danke ich f\"ur wertvolle Hinweise zur Gestaltung der Frageb\"ogen und J.~Mordel f\"ur Anregungen zur Gestaltung des Lehrforschungsprojektes.

\end{document}